\documentstyle[epsfig]{mn}

\title[]{UV/Optical Emission Accompanying Gamma-ray Burst}

\author[]{Y. Z. Fan$^{1,2 \star}$ and D. M. Wei$^{1,2}$
\thanks{E-mail: yzfan@pmo.ac.cn(YZF); dmwei@pmo.ac.cn(DMW)} \\
$^1${\sl Purple Mountain Observatory, Chinese Academy of
Science, Nanjing 210008, China}\\
       $^2${\sl National Astronomical
Observatories, Chinese Academy of Sciences, Beijing 100012,
China}}

\date{Accepted ......  Received ......; in original form ......}

\begin{document}
\voffset=-0.5 in

\maketitle
\begin{abstract}
We discuss the possible simultaneously UV/optical emission
accompanying Gamma-ray bursts (GRBs). We show that as long as the
intrinsic spectrum of GRB can extend to $\sim$10 GeV or higher,
there is a large amount of relativistic $e^\pm$ pairs generated
due to the annihilation of the soft $\gamma-$rays with the very
energetic photons, which dominates over the electrons/positrons
associated with the fireball, no matter the fireball is highly
magnetized or not (For the highly magnetized fireball, the
magnetic field is ordered, the high linear polarization of the
multi-wavelength emission is expected). We find that these $e^\pm$
pairs can power an UV flash with $m\simeq 12-13{\rm th}$
magnitude, and the corresponding optical emission can be up to
$m_{\rm R}\simeq15-16{\rm th}$ magnitude. Such bright UV emission
can be detected by the upcoming satellite Swift, planned for
launch in 2004. The behavior of the optical-UV spectrum
($F_{\nu}\propto \nu^{5/2}$) differs significantly from that of
the reverse shock emission ($F_{\nu}\propto \nu^{-\beta/2}$,
$\beta \simeq 2.2$), which is a signature of the emission
accompanying with GRB. The mild optical emission can be detected
with the ROTSE-IIIa telescope system, if the response to the GRB
alert is fast enough.

\end{abstract}
\begin{keywords}
Gamma-rays: bursts---radiation mechanism: non-thermal
\end{keywords}

\section{Introduction}
Although the central engine for Gamma-ray bursts (GRBs) is far
from clear, it is generally suggested that $\gamma$-ray bursts are
powered by the dissipation of energy in a highly relativistic
wind, driven by gravitational collapse of massive star into a
neutron star or a black hole (see M\'{e}sz\'{a}ros 2002 and Cheng
\& Lu 2001 for recent reviews). There are two possible models
involving this scenario: One is the internal shocks model
(Paczy\'{n}ski \& Xu 1994; Rees \& M\'{e}sz\'{a}ros 1994)
involving an baryon-rich fireball, which can reproduce the
observed temporal structure in GRBs naturally. The other is the
Poynting flux driven outflows from magnetized rotators (Usov 1994;
Thompson 1994; M\'{e}sz\'{a}ros \& Rees 1997). Comparing with the
widely accepted baryon-rich fireball model, the Poynting flux
model is of more and more interest, since: (1) It provides us the
most plausible explanation for the very high linear polarization
during the prompt $\gamma-$ray emission of GRB 021206 (e.g. Coburn
\& Boggs 2003; Lyutikov, Pariev \& Blandford 2003), although some
alternative explanations still remain (e.g. Shaviv \& Dar 1995;
Waxman 2003). (2) Modeling the reverse shock emission of GRB
990123 indicates that the reverse shock region should anchor a
strong field far more than the forward shock region holds (Fan et
al. 2002; Zhang, Kobayashi \& M\'{e}sz\'{a}ros 2003), which hints
the fireball may be highly magnetized (see the review of Zhang \&
M\'{e}sz\'{a}ros (2003) for more detail).

In the afterglow epoch, there is few theoretical attentions have
been paid on the emission at UV/optical emission during
$\gamma-$ray burst phase. Pilla \& Loeb (1998) have discussed such
emission in the internal shocks. However, in their work, the
synchrotron self-absorption effect has been ignored (see their
figure 3, the spectrum at the optical-UV band takes the form of
$F_{\rm \nu}\propto \nu^{1/3}$). Conversely, before the discovery
of the afterglow, this topic is of some interest (e.g., Schaefer
et al. 1994; Katz 1994; Wei \& Cheng 1997).  In this paper, we
reinvestigate that topic in some detail. Different from the
previous works, here we emphasize the contribution of the $e^\pm$
pairs generated in the phase of $\gamma-$burst---as long as the
intrinsic spectrum can extend to $\sim$10GeV or higher, there is a
large amount of relativistic $e^\pm$ pairs generated due to the
annihilation of the soft $\gamma-$rays with the energetic photons
with energy up to 10GeV, just as realized by several authors
previously (e.g. Pilla \& Leob 1998; Guetta, Spada \& Waxman 2001;
M\'{e}sz\'{a}ros et al. 2002; Li et al. 2003; Fan, Dai \& Lu
2004)\footnote{During revising our paper, a great detailed
numerical calculation work appeared (Pe'er \& Waxman 2003).}.
There are some evidences that the spectra of some GRBs extend to
$>100$ MeV (e.g. Schaefer et al. 1998; Gonz\'{a}lez et al. 2003;
Guetta \& Granot 2003). For some GRBs observed by EGRET, a
significant fraction of power has been emitted into the GeV energy
range, and the spectra can be described by a single power-law
ranging from MeV to GeV energy band (See Fishman \& Meegan 1995
for a review). We find that those resulting $e^\pm$ pairs power a
bright UV flash with $m\simeq12-13{\rm th}$ magnitude, the
corresponding optical emission can be up to $m_{\rm
R}\simeq15-16{\rm th}$ magnitude. Such bright UV emission can be
detected by the upcoming Satellite Swift, planned for launch in
2004. The mild optical emission can be detected with the
ROTSE-IIIa telescope system, if the response to the GRB alert is
fast enough.

This paper is structured as follows: In section 2, we discuss the
pair-production and the possibility of annihilation in
$\gamma-$ray burst phase. In section 3, we calculate the
synchrotron radiation of these new born relativistic $e^\pm$
pairs. Conclusions and some discussions on observation are given
in section 4.

\section{Pair Loading in GRBs}
History, a large Lorentz factor $\eta \sim 100$$-$$1000$ is first
introduced to avoid the ``compactness problem'' in GRBs. However,
a significant fraction of photons with very energetic energy may
still suffer from absorption and yielding considerable amount of
$e^\pm$ pairs. Pilla \& Leob (1998) have studied spectral
signatures of GRB itself by considering pair production and shown
there are a large amount of $e^\pm$ pairs generated. The co-moving
annihilation time of these pairs is longer than the hydrodynamical
time, so they can survive in the wind for a long time. For this
reason, Li et al. (2003) have reinvestigated the pair generation
in the $\gamma-$ray burst phase and studied the reverse shock
emission powered by a such pair-rich fireball interacting with the
interstellar medium (ISM), and shown there comes the very strong
IR flashes. In those two works, the resulting pairs dominate over
the electrons associated with baryons. Generally, the typical
thermal Lorentz factor of these resulting $e^\pm$ pairs is only
several tens, its synchrotron radiation contributes little to the
$\gamma-$ray band. Consequently, only its inverse Compton (IC)
radiation has been calculated (Pilla \& Loeb 1998; Fan, Dai \& Lu
2004). In this paper we turn to investigate the emission at the
much lower energy band, i.e., the UV/optical.

As mentioned early, for the most energetic photons at the high
energy end of the spectrum, the optical depth of $\gamma-\gamma$
absorption may exceed unity. As a result, rather than escaping
from the outflow, these photons are absorbed by the soft
$\gamma-$rays and yielding relativistic $e^\pm$ pairs. Below,
following Li et al. (2003) and Fan, Dai \& Lu (2004), we calculate
how many pairs are generated in this process for a pulse with a
typical variable timescale $\delta t\sim 0.1{\rm s}$ (Shen \& Song
2003).

An excellent phenomenological fit for the GRB spectrum was
introduced by Band et al. (1993), which is characterized by two
power laws joined smoothly at a break frequency $\nu_{\rm
b}\approx 1.21\times 10^{20}~{\rm Hz}$. For $\nu>\nu_{\rm b}$, the
photon spectra can be approximated by a power law $dN/d\nu=N_{\rm
\nu_{\rm b}}(\nu/\nu_{\rm b})^{-(p+2)/2}$, where $N_{\rm \nu_{\rm
b}}=(p-2)(h\nu_{\rm b}^2)^{-1}L\delta t/2(p-1)$ (Dai \& Lu 2002),
$p\simeq2.5$ is the index for a power-law distribution of
relativistic electrons/positrons account for the observed
$\gamma-$ray emission. Therefore for $\nu>\nu_{\rm b}$ we have
\begin{eqnarray}
N_{\rm >\nu}&=&\int_{\rm \nu}^{\infty} (dN/d\nu) d\nu\nonumber\\
&=&[(p-2)/p(p-1)](h\nu_{\rm b})^{-1}(\nu/\nu_{\rm
b})^{-p/2}L\delta t.
\end{eqnarray}
A photon with energy $h\nu$ may annihilate any photons above the
energy $h\nu_{\rm an}=(\eta m_{\rm e}c^2)^2/h\nu$, the optical
depth is given by (Lithwick \& Sari 2001; Dai \& Lu 2002; Li et
al. 2003)
\begin{equation}
\tau_{\rm \gamma \gamma}(\nu)={(11/180)\sigma_{\rm T}N_{\rm
>\nu_{\rm an}}\over 4\pi (\eta^2 c\delta t)^2}
\end{equation}
A photon with $\tau_{\rm \gamma \gamma}(\nu)>1$ would be absorbed
then deposited in the fireball. The condition $\tau_{\rm \gamma
\gamma}(\nu)=1$ results in $\nu_{\rm an}\approx 6.4\times
10^{20}~{\rm Hz}~\nu_{\rm b,20.1}^{(p-2)/p}L_{52}^{2/p}\delta
t_{-1}^{-2/p}\eta_{2.5}^{-8/p}$ (In this paper, we adopt the
convention $Q_{\rm x}=Q/10^{\rm x}$ for expressing the physical
parameters, using cgs units). The cut off frequency is
\begin{equation}
\nu_{\rm cut}=2.2\times 10^{24}~{\rm Hz}~\nu_{\rm
b,20.1}^{(2-p)/p}L_{\rm 52}^{-2/p}\delta t_{-1}^{\rm
2/p}\eta_{2.5}^{(2p+8)/p},
\end{equation}
the corresponding ``thermal'' Lorentz factor of resulting $e^\pm$
pairs reads
\begin{equation}
\gamma_{\rm pair,m}\approx{h\nu_{\rm cut}\over 2\eta m_{\rm
e}c^2}=29\nu_{\rm b,20.1}^{(2-p)/p}L_{52}^{-2/p}\delta
t_{-1}^{2/p}\eta_{2.5}^{(p+8)/p}.
\end{equation}
The total number of the resulting $e^\pm$ pairs is
\begin{eqnarray}
N_{\rm e^\pm}&=&[(p-2)/p(p-1)](h\nu_{\rm
b})^{-1}(\nu_{\rm cut}/\nu_{\rm b})^{-p/2}L\delta t\nonumber\\
&=&7.4\times 10^{50}L_{52}^2\eta_{2.5}^{-(p+4)}\nu_{\rm
b,20.1}^{p-2}.
\end{eqnarray}

In principle, more detailed numerical calculation is needed to
calculate the number of generating $e^\pm$ pairs just as Pilla \&
Loeb (1998) and Guetta et al. (2001) have done. However, here
we'll show that our analytical estimation (equation (5)) coincides
with the numerical results of Pilla \& Loeb (1998) quite well. For
the numerical example holding in Pilla \& Loeb (1998): $E\approx
10^{51}~{\rm ergs}$, $M\approx 10^{27}~{\rm g}$, $\delta t\approx
0.01~{\rm s}$ and $\eta\approx 400$ (where $M$ is the mass of the
shell, and other parameters mean as usual). With these values, our
equation (5) reads $N'_{\rm e^{\pm}}\approx 1.14\times
10^{52}L_{53}^2\eta_{2.6}^{\rm -(p+4)}\nu_{\rm b,20.1}^{\rm p-2}$.
On the other hand, the number of the electron associated with
baryons $N'_{\rm e}=M/m_{\rm p}\approx 6.0\times 10^{50}$. The
ratio of them is $2N'_{\rm e^{\pm}}/N'_{\rm e}\approx 38$. Please
note that in the original work of Pilla \& Loeb (see the caption
of figure 1 (b), astro-ph/97120219), they obtain a ratio of about
40! Such excellent coincidence implies our analytical treatment is
reasonable and our results are reliable.

These resulting $e^\pm$ pairs may annihilate each other into
$\gamma$-ray again. That possibility has been discussed in great
detail by Pilla \& Loeb (1998), Li et al. (2003) and Fan, Dai \&
Lu (2004). For typical GRB parameters, $\eta\sim 300$, $\delta
t\sim 0.1{\rm s}$, $L\sim 10^{52}{\rm ergs~s^{-1}}$, the
annihilation time of these pairs is much longer than the
hydrodynamic time (measured in the comoving frame). As a result,
these new born $e^\pm$ pairs survive in the wind for long time
rather than annihilate locally, only for which are our following
calculations valid.

\section{Synchrotron Emission at the UV/Optical Band}
The resulting $e^\pm$ pairs are in fast cooling phase, most of
them cool down to $\gamma_{\rm e}\sim 1$ rapidly. However, it is
well known that only the emission of the relativistic
electrons/positrons can be described by the synchrotron radiation.
For the sub-relativistic electrons, their cyclotron radiation
contribute little to the UV/optical emission which is of our
interest. Therefore, in this paper, only the electrons with the
Lorentz factor $\gamma_{\rm e}>\gamma_{\rm e,c}$ ($\gamma_{\rm
e,c}$ represents a critical Lorentz factor, which plays the same
role as the ``cooling Lorentz factor'' elsewhere) have been taken
into account. As an illustration, we take $\gamma_{\rm e,c}=2$. In
the following discussion, the superscripts ``$m$'' and ``$b$''
represent the highly magnetized fireball and baryon-rich fireball
respectively.

\subsection{Baryon-rich Fireball}
For a baryon-rich fireball, the number of the total electrons
account for the observed $\gamma-$ray emission can be estimated by
$N_{\rm e,tot}\eta \gamma_{\rm m}m_{\rm e}c^2=E_{\rm \gamma}$,
$\gamma_{\rm m}$ is the typical Lorentz factor of electrons, which
is constrained by $\eta \gamma_{\rm m}^2 eB^{\rm b}/2\pi m_{\rm
e}c=\nu_{\rm b}$. $B^{\rm b}$ is the strength of the co-moving
magnetic field, which can be estimated as follows: Assuming
$E_{\rm B}=\epsilon_{\rm B}E_{\rm \gamma}$ ($E_{\rm B}$ being the
total magnetic energy carried by the fireball, $E_{\rm \gamma}$
being the total energy emitted in the $\gamma-$ray band), we have
$4\pi (\eta^2 c\delta t)^2 c ({B^{\rm
b}}^2/8\pi)\eta^2=\epsilon_{\rm B}L$, which leads to $ B^{\rm
b}=2.9\times 10^3~{\rm Guass}~L_{52}^{1/2}\eta_{2.5}^{-3}\delta
t_{-1}^{-1}\epsilon_{\rm B,-1}^{1/2}$. After some simple algebra
we have
\begin{equation}
N_{\rm e,tot}=5.8\times 10^{52}E_{\rm
\gamma,53}L_{52}^{1/4}\eta_{2.5}^{-2}\delta
t_{-1}^{-1/2}\epsilon_{\rm B,-1}^{1/4}\nu_{\rm b,20.1}^{-1/2}.
\end{equation}
For a typical pulse with typical variable timescale $\delta t$,
the electrons contained can be estimated as
\begin{equation}
N_{\rm e}=N_{\rm e,tot}\delta t/T=5.8\times 10^{50}L_{\rm
52}^{5/4}\eta_{2.5}^{-2}\delta t_{-1}^{1/2}\epsilon_{\rm
B,-1}^{1/4}\nu_{\rm b,20.1}^{-1/2}.
\end{equation}
where $T\sim 10~{\rm s}$ (measured in the local frame) is the
typical ``effective'' duration of GRBs, within which the
lightcurve is relatively smooth and most of the total energy has
been emitted. Thus $L\approx E_{\rm \gamma}/ T$. Now we can define
a coefficient $k_\pm$ as
\begin{equation}
k_{\pm}\equiv N_{\rm e^\pm}/N_{\rm e}
=1.3L_{52}^{3/4}\eta_{2.5}^{-(p+2)}\nu_{\rm b,20.1}^{p-3/2}\delta
t_{-1}^{-1/2}\epsilon_{\rm B,-1}^{-1/4}.
\end{equation}

With equation (4), the characteristic synchrotron emission
frequency of the new born $e^\pm$ can be estimated by
\begin{eqnarray}
\nu_{\rm m}^{\rm b}&=&\eta \gamma_{\rm pair,m}^2{eB^{\rm b}\over
2\pi m_{\rm e}c}\nonumber\\
&\approx & 2.0\times 10^{15}~{\rm Hz}~\epsilon_{\rm B,-1}^{1\over
2}\nu_{\rm b,20.1}^{2(2-p)\over p}L_{52}^{(p-8)\over 2p}\delta
t_{\rm -1}^{(4-p)\over p}\eta_{2.5}^{16\over p}.
\end{eqnarray}
where $m_{\rm e}$ being the mass of the electron.

The electron with a Lorentz factor $\gamma_{\rm e}$
($\gg\gamma_{\rm e,c}$) cools down to $\gamma_{\rm e,c}$ at a
timescale
\begin{equation}
t_{\rm life}^{\rm b}\approx3\pi m_{\rm e}c/\sigma_{\rm T}{B^{\rm b
}}^2\eta\gamma_{\rm e,c}\approx0.076~{\rm s}~\epsilon_{\rm B,-1
}^{-1}L_{52}^{-1}\eta_{2.5}^5\delta t_{-1}^{2}.
\end{equation}
which is comparable with $\delta t$. The characteristic frequency
with respect to $\gamma_{\rm e,c}$ is
\begin{equation}
\nu_{\rm c}^{\rm b}=\eta \gamma_{\rm e,c}^2{eB^{\rm b}\over 2\pi
m_{\rm e}c}\approx 9.8\times 10^{12}~{\rm Hz}~\epsilon_{\rm
B,-1}^{1/2}L_{\rm 52}^{1/2}\eta_{2.5}^{-2}\delta t_{-1}^{-1}.
\end{equation}

To calculate the observed energy flux, the synchrotron
self-absorption effect must be considered. Now, the
electrons/positrons can be classed into two components: one is the
electrons account for the $\gamma-$ray emission with the
distribution $dn/d\gamma_{\rm e}\propto \gamma_{\rm e}^{-2}$ for
$\gamma_{\rm e,c}<\gamma_{\rm e}<\gamma_{\rm m}$ and
$dn/d\gamma_{\rm e}\propto \gamma_{\rm e}^{-(p+1)}$ for
$\gamma_{\rm e}>\gamma_{\rm m}$. The other is the resulting
$e^\pm$ pairs with the distribution $dn/d\gamma_{\rm e}\propto
\gamma_{\rm e}^{-2}$ for $\gamma_{\rm e,c}<\gamma_{\rm
e}<\gamma_{\rm pair,m}$ and $dn/d\gamma_{\rm e}\propto \gamma_{\rm
e}^{-(p+4)/2}$ for $\gamma_{\rm e}>\gamma_{\rm pair,m}$. For the
former, the synchrotron self-absorption frequency can be estimated
by (see the Appendix of Wu et al. 2003 for detail): $\nu_{\rm
a}=2.4\times 10^{15}~{\rm Hz}~L_{52}^{3/4}\eta_{2.5}^{-3}\delta
t_{-1}^{-7/6}\epsilon_{\rm B,-1}^{5/12}\nu_{\rm b,20.1}^{-1/6}$.
For the latter, the synchrotron self-absorption frequency can be
estimated by (Wu et al. 2003): $\nu_{\rm a}'=4.0\times 10^{15}{\rm
Hz}[\epsilon_{\rm B,-1}^{p/2}\eta_{2.5}^{-4(p+2)}\delta
t_{-1}^{-(p+8)}L_{52}^{\rm (8+p)/2}\nu_{\rm
b,20.1}^{2(p-2)}]^{1/(p+12)}$. In practice, the self-absorption
frequency of such a system is determined by these two components
corporately rather than separately. Here, as a rough estimation,
we assume the actual self-absorption frequency is about $k~(\simeq
1+O(0.1))$ times $\nu_{\rm a}'$, therefore
\begin{eqnarray}
\nu_{\rm a}^{\rm b}&=&4.0\times 10^{15}{\rm Hz}~k[\epsilon_{\rm
B,-1 }^{p/2}\eta_{2.5}^{-4(p+2)}\nonumber\\
&& \delta t_{-1}^{-(p+8)}L_{52}^{\rm (8+p)/2}\nu_{\rm
b,20.1}^{2(p-2)}]^{1/(p+12)}
\end{eqnarray}

Please note that $\nu_{\rm m}^{\rm b}$, $\nu_{\rm c}^{\rm b}$ and
$\nu_{\rm a}^{\rm b}$ are all measured in the local frame. As
translating into the observer frame, all of them need to be
multiplied by $1/(1+z)$, $z\sim 1$ being the typical redshift of
GRBs. The peak flux can be estimated by
\begin{eqnarray}
F_{\rm \nu_{\rm a}}^{\rm b}&=&F_{\rm \nu_{\rm max}}^{\rm b
}({\nu_{\rm m}^{\rm b}\over \nu_{\rm c}^{\rm b}})^{-1/2}({\nu_{\rm
a}^{\rm b }\over \nu_{\rm
m}^{\rm b}})^{-(p+2)/4}\nonumber\\
&=&0.144~{\rm Jy}~\epsilon_{\rm B,-1}^{p-3\over
p+12}k^{-(p+2)/4}({1+1/2k_\pm\over
1.38})\eta_{2.5}^{2(14-3p)\over p+12}\nonumber\\
&& \delta t_{-1}^{10\over p+12}L_{52}^{p+7\over p+12}\nu_{\rm
b}^{5(p-2)\over p+12}({1+z\over 2})D_{\rm L,28.34}^{-2}.
\end{eqnarray}
where $F_{\rm \nu_{\rm max}}^{\rm b}=N_{\rm rad}^{\rm b}\eta
P_{\rm \nu_{\rm m}}^{\rm b}(1+z)/4\pi D_{\rm L}^2$, $N_{\rm
rad}^{\rm b}=2N_{\rm e^\pm}(1+{1\over 2k_\pm})t_{\rm life}^{\rm b
}/\delta t$, $P_{\rm \nu_{\rm m}}^{\rm b}=e^3B^{\rm b}/m_{\rm
e}c^2$. $D_{\rm L}$ is the luminosity distance of the source (we
take $H_0=65{\rm km~s^{-1}~MpC^{-1}}$, $\Omega_{\rm m}=0.3$,
$\Omega_\wedge=0.7$).

Since $\nu_{\rm c}^{\rm b}<\nu_{\rm R,obs}=5\times 10^{14}{\rm
Hz}<\nu_{\rm m}^{\rm b}<\nu_{\rm a}^{\rm b}$, the observed R band
flux can be estimated by
\begin{eqnarray}
F_{\rm \nu_{\rm R,obs}}^{\rm b}&=&F_{\rm \nu_{\rm a}}^{\rm b}
[{(1+z)\nu_{\rm R,obs}\over \nu_{\rm a}^{\rm b}}]^{5/2}\nonumber\\
&=&4.5~{\rm mJy}~k^{-(p+12)/4}[(1+1/2k_\pm)/1.38]\nonumber\\
&&({1+z\over 2})^{7/2}D_{\rm L,28.34}^{-2}\epsilon_{\rm
B,-1}^{-1/4}L_{\rm 52}^{-1/4}\eta_{2.5}^{4}\delta t_{-1}^{5/2}.
\end{eqnarray}
which hints for typical parameters, the optical emission is weak to $m_{\rm
R}\approx 15$.

Similarly, for $\nu_{\rm obs}=1.8\times 10^{15}~{\rm Hz}$
corresponding to the wavelength $\lambda=170$ nm, the upper limit
of UVOT carried by Swift (see in http://swift.gsfc.nasa.gov), the
predicted emission is high up to $m\simeq 12{\rm th}$ magnitude.

Pe'er \& Waxman (2003) have calculated the prompt GRB spectra
($>0.1{\rm keV}$) in great detail within the fireball model
framework, and some important effects such as the $e^\pm$ pairs
production/annihilation and so on have been taken into account. In
order to estimate the validity of our calculation, we compare our
results with the detailed numerical calculation (Pe'er \& Waxman
2003), and find that our results do not show much difference from
theirs. For example, for their low compactness case shown in their
figure 4: $L=10^{52}{\rm ergs}$, $\epsilon_{\rm e}=\epsilon_{\rm
B}=10^{-0.5}$, $p=3$, $\delta t=0.01~{\rm s}$ and $\eta=300$. The
flux $F_{\nu_{\rm R,obs}}\sim 5\times 10^{-5}{\rm Jy}$ (we have
extended their figure to $h\nu_{\rm R,obs}\approx 2$ eV energy
range, at which energy band $F_{\nu}\propto \nu^{5/2}$). With
these parameters, our simple analytic result (see Eq. (14)) gives
$F_{\nu_{\rm R,obs}}\sim 1.4\times 10^{-5}{\rm Jy}$. Therefore, as
an approximation, we think our analytic results can be used to
estimate the UV/Optical emission from GRBs.

Below we discuss the possible SSC (synchrotron self-Compton)
radiation briefly. The typical energy of the SSC radiation can be
estimated by $h\nu_{\rm m}^{\rm SSC}\simeq 2\gamma_{\rm pair,m}^2
h\nu_{\rm m}\approx 18~{\rm keV}$. The ratio of the SSC luminosity
($L_{\rm SSC}$) to the synchrotron luminosity ($L_{\rm syn}$) of
$e^\pm$ pairs can be estimated by $x\equiv{L_{\rm SSC}\over L_{\rm
syn}}={U_{\rm e}/(1+x)\over U_{\rm B}^{\rm b}}$, where $U_{\rm
e}/U_{\rm B}^{\rm b}$ are the electron/magnetic energy density
respectively. Hence $x=(-1+\sqrt{1+4U_{\rm e}/U_{\rm B}^{\rm
b}})/2=(-1+\sqrt{1+4 E_{\rm \gamma,\nu>\nu_{\rm
cut}}/\epsilon_{\rm B} E_{\rm \gamma}})/2\sim 0.6$ for
$\epsilon_{\rm B}\simeq 0.1$. $L_{\rm syn}/L\approx E_{\rm
\gamma,\nu>\nu_{\rm cut}}/(1+x)E_{\rm \gamma}\approx (\nu_{\rm
cut}/\nu_{\rm b})^{\rm (2-p)/2}/(1+x)$. Therefore $L_{\rm
SSC}/L\approx x (\nu_{\rm cut}/\nu_{\rm b})^{\rm
(2-p)/2}/(1+x)\approx 0.04$. So the SSC component can change the
observed soft $\gamma-$ray spectrum in some degree, which may help
to explain the observed X-ray excess in some GRBs (Band et al.
1993).

\subsection{Highly Magnetized Fireball}
For the highly magnetized fireball, the characteristic synchrotron
emission frequency of the new born $e^\pm$ pairs can be estimated
by
\begin{eqnarray}
\nu_{\rm m}^{\rm m}&=&\eta \gamma_{\rm pair,m}^2{eB^{\rm m}\over
2\pi m_{\rm e}c}\nonumber\\
&\approx & 4.4\times 10^{15}~{\rm Hz}~\epsilon^{1\over 2}\nu_{\rm
b,20.1}^{2(2-p)\over p}L_{52}^{(p-8)\over 2p}\delta t_{\rm
-1}^{(4-p)\over p}\eta_{2.5}^{16\over p}.
\end{eqnarray}
At the present case, we assume the electromagnetic energy
dominated over other ones, thus $B^{\rm m}$ can be estimated by
$B^{\rm m}\approx6.3\times10^3~{\rm Gauss}~\epsilon^{1/2}L_{\rm
52}^{1/2}\eta_{2.5}^{-3}\delta t_{-1}^{-1}$, $\epsilon=E_{\rm
B}/E_{\rm \gamma}\sim 1$.

Now, the electron with a Lorentz factor $\gamma_{\rm e}$
($\gg\gamma_{\rm e,c}$) cools down to $\gamma_{\rm e,c}$ at a
timescale
\begin{equation}
t_{\rm life}^{\rm m}\approx3\pi m_{\rm e}c/\sigma_{\rm T}{B^{\rm
m}}^2\eta\gamma_{\rm e,c}\approx0.016~{\rm
s}~\epsilon^{-1}L_{52}^{-1}\eta_{2.5}^5\delta t_{-1}^{2}.
\end{equation}
The characteristic frequency with respect to $\gamma_{\rm e,c}$ is
\begin{equation}
\nu_{\rm c}^{\rm m}=\eta \gamma_{\rm e,c}^2{eB^{\rm m}\over 2\pi
m_{\rm e}c}\approx 2.1\times 10^{13}~{\rm Hz}~\epsilon^{1/2}L_{\rm
52}^{1/2}\eta_{2.5}^{-2}\delta t_{-1}^{-1}.
\end{equation}
The synchrotron self-absorption frequency ($\nu_{\rm a}>\nu_{\rm
m}$) can be estimated by (Wu et al. 2003)
\begin{eqnarray}
\nu_{\rm a}^{\rm m}&=&5.6\times 10^{15}{\rm
Hz}[\epsilon^{p/2}\eta_{2.5}^{-4(p+2)}\delta
t_{-1}^{-(p+8)}\nonumber\\
&&L_{52}^{\rm (8+p)/2}\nu_{\rm b,20.1}^{2(p-2)}]^{1/(p+12)}.
\end{eqnarray}
where it is assumed that the electrons/positrons carried by the
fireball is far less than the $e^\pm$ pairs generated in the
$\gamma-$ray phase (Zhang \& M\'{e}sz\'{a}ros 2002). Now the peak
flux can be estimated by
\begin{eqnarray}
F_{\rm \nu_{\rm a}}^{\rm m}&=&F_{\rm \nu_{\rm max}}^{\rm
m}({\nu_{\rm m}^{\rm m}\over \nu_{\rm c}^{\rm
m}})^{-1/2}({\nu_{\rm a}^{\rm m}\over \nu_{\rm
m}^{\rm m}})^{-(p+2)/4}\nonumber\\
&=&0.096~{\rm Jy}~\epsilon^{p-3\over
p+12}\eta_{2.5}^{2(14-3p)\over p+12}\delta t_{-1}^{10\over
p+12}\nonumber\\
&&L_{52}^{p+7\over p+12}\nu_{\rm b}^{5(p-2)\over p+12}({1+z\over
2})D_{\rm L,28.34}^{-2}.
\end{eqnarray}
where $F_{\rm \nu_{\rm max}}^{\rm m}=N_{\rm rad}^{\rm m}\eta
P_{\rm \nu_{\rm m}}^{\rm m}(1+z)/4\pi D_{\rm L}^2$, $N_{\rm
rad}^{\rm m}=2N_{\rm e^\pm}t_{\rm life}^{\rm m}/\delta t$, $P_{\rm
\nu_{\rm m}}^{\rm m}=e^3B^{\rm m}/m_{\rm e}c^2$.

The observed R band flux can be estimated by
\begin{eqnarray}
F_{\rm \nu_{\rm R,obs}}^{\rm m}&=&F_{\rm \nu_{\rm a}}^{\rm m}
[{(1+z)\nu_{\rm R,obs}\over \nu_{\rm a}^{\rm m}}]^{5/2}\nonumber\\
&=&1.4~{\rm mJy}~({1+z\over 2})^{7/2}D_{\rm
L,28.34}^{-2}\epsilon^{-1/4}\nonumber\\
&&L_{\rm 52}^{-1/4}\eta_{2.5}^{4}\delta t_{-1}^{5/2}.
\end{eqnarray}
which hints for typical parameters, the optical emission is weak
to $m_{\rm R}\approx 16$.

Similarly, for $\nu_{\rm obs}=1.8\times 10^{15}~{\rm Hz}$, the
predicted emission is up to $m\simeq 13{\rm th}$ magnitude.

In the present case, the typical energy of the SSC radiation can
be estimated by $h\nu_{\rm m}^{\rm SSC}\simeq 2\gamma_{\rm
pair,m}^2 h\nu_{\rm m}\approx 35~{\rm keV}$. Now,
$x=(-1+\sqrt{1+4U_{\rm e}/U_{\rm B}^{\rm m}})/2=(-1+\sqrt{1+4
E_{\rm \gamma,\nu>\nu_{\rm cut}}/\epsilon E_{\rm
\gamma}})/2\approx E_{\rm \gamma,\nu>\nu_{\rm cut}}/\epsilon
E_{\gamma}$ for $E_{\rm \gamma,\nu>\nu_{\rm cut}}\ll \epsilon
E_{\rm \gamma}$. Thus $L_{\rm SSC}/L\approx 1/[(1+x)\epsilon
(\nu_{\rm cut}/\nu_{\rm b})^{\rm (p-2)}]\approx
0.01\epsilon^{-1}$, which implies that the SSC component can not
change the observed soft $\gamma-$ray spectrum significantly, at
least for the typical parameters taken here.

\section{Discussion and Conclusions}
GRBs are characterized by emission in the few hundred $\rm keV$
ranges with a non-thermal spectrum, X-ray emission is
weaker$-$only a few percent of the energy is emitted below 10 keV
and prompt emission at lower energies has not been observed so
far. One exception is the optical flash accompanying with GRB
990123 (Akerlof et al. 1999), which is believed to be powered by
the reverse shock (Sari \& Piran 1999). If such emission is the
low-energy tail of the $\gamma-$ray emission, the light curves in
the different energy bands should be highly correlated, which is
not the case (Sari \& Piran 1999). Akerlof et al. (2000) have
performed a search for optical counterparts to six GRBs with
location errors of 1 square degree or better, but no optical
counterpart has been detected, the earliest limiting sensitivity
is $m>13.1$ at 10.85 seconds after the gamma-ray rises. No
simultaneous optical emission from GRBs has been detected by Kehoe
et al. (2001), too. All of these observations suggest that the
simultaneous optical emission should not be typically brighter
than 14th magnitude, which coincides with our results presented
here.

The typical prompt optical emission predicted in this paper,
$m_{\rm R}\sim 15-16$th magnitude, is significantly stronger than
$m_{\rm v}\sim 18$th magnitude predicted by Katz (1994). Such
emission can be detected by the current ROTSE-IIIa telescope
system, which is a 0.45-m robotic reflecting telescope and managed
by a fully-automated system of interacting daemons within a Linux
environment. The telescope has an f-ratio of 1.9, yielding a field
of view of 1.8${\rm \times}$1.8 degrees. The control system is
connected via a TCP/IP socket to the Gamma-ray Burst Coordinate
Network (GCN), which can respond to GRB alerts fast enough
($<10{\rm s}$). ROSTE-IIIa can reach 17th magnitude in a 5-s
exposure, 17.5 in 20-s exposure (see Smith et al. 2003 for
detail), which is sufficient to detect the optical emission
predicted in this paper. However, for the standard fireball model,
the very early optical emission powered by the reverse shock can
be up to $m_{\rm R}\sim 9$th magnitude or even brighter (Sari \&
Piran 1999; Wu et al. 2003; Li et al. 2003), which far surpass the
optical emission predicted here. In practice, such strong early
optical emission should be very rare, since it has not been
detected for most GRBs (Akerlof et al 2000; Kehoe et al. 2001). It
is unclear why the early optical emission is so weak. If the
fireball is highly magnetized, such emission may be weak to
$m_{\rm R}\sim 14$th magnitude at the deceleration radius $R_{\rm
dec}\sim 10^{17}~{\rm cm}$, the corresponding timescale $t_{\rm
obs}\sim 20~{\rm s}~(1+z)R_{\rm dec, 17}\eta_{2.5}^{-2}$ (Fan, Wei
\& Wang 2004). In the collision model of the magnetized wind and
the external medium proposed by Smolsky \& Usov (2000), the
synchrotron radiation generated in the vicinity of the wind front
can be high up to tens of MeV, rather than eV as we generally
suggested. In this case, the very early optical afterglow is very
weak. So, the optical emission accompanying with GRB may be
detected independently.

Due to the strong synchrotron self-absorption, the emission peaks
at UV band. For $\nu_{\rm obs}=1.8\times 10^{15}{\rm Hz}$
($\lambda=170~{\rm nm}$), the typically simultaneous emission is
high up to $m\simeq 12-13{\rm th}$ magnitude, which is bright
enough to be detected by the UVOT (covering $170~{\rm nm}-650~{\rm
nm}$ with 6 colors) carried by Swift, planned for launch in early
2004. The observation of that UV emission is important, since: at
UV band, the spectrum predicted in this paper takes the form of
$F_{\rm \nu}\propto \nu^{5/2}$, which is significantly different
from that of the reverse shock emission, $F_{\rm \nu}\propto
\nu^{-\beta/2}$, where $\beta\simeq2.2$ is the index of the
power-law distribution of the relativistic electrons heated by the
reverse shock (Sari \& Piran 1999; Fan et al. 2002; Wu et al.
2003). The flux of UV emission predicted here is far above that of
the optical emission, it is quite the contrary for the reverse
shock emission. Therefore, the spectrum feature at the optical-UV
band ($F_{\nu}\propto \nu^{5/2}$) is a signature of the emission
accompanying GRBs.

For both the baryon-rich fireball and highly magnetized one, the
generated $e^\pm$ pairs dominated over the electrons (including
positrons) associated with the fireball. For reproducing the
observed $\gamma-$ray emission, the magnetic field strengthes for
these two type fireballs are comparable. Consequently, the
UV/optical emission predicted here do not show much difference for
these two kinds of fireballs. However, for the highly magnetized
fireball, the magnetic field is ordered, so the high linear
polarization of the synchrotron radiation at multi-wavelength
bands is expected.

Pilla \& Loeb (1998) have discussed the possible IC scattering of
the resulting $e^\pm$ pairs with the intrinsic GRB photons, and
found that if $U_{\rm \gamma}\gg U_{\rm m}$, the pairs transfer
nearly all of their energy back to the radiation field via IC
scattering. Fortunately, for the two cases discussed here, $U_{\rm
\gamma}$ and $U_{\rm m}$ are comparable, so that IC process may be
important but is not dominant, especially for the highly
magnetized fireball. Therefore it will not change our result
presented here significantly (The detailed numerical research is
beyond the scope of this Paper).

It should be noted that in this paper our results are based on the
simple analytic analysis, which is a great simplification of the
real situation. In our calculation, some important effects such as
the pair annihilation have not been taken into account. However,
as described in previous section, we found that our results are
not much different from those of detailed numerical calculations
(e.g., Pilla $\&$ Leob 1998; pe'er $\&$ waxman 2003). This
suggests that as an order estimation, our present work is
reliable. Furthermore, our work has the benefit of showing
scalings with parameters better. We have shown that the predicted
flux in the UV/Optical band accompanying GRBs strongly depends on
the typical variability timescale ($\delta t$) as well as the
typical bulk Lorentz factor ($\eta$), i.e., $F \propto \eta^4
\delta t^{5/2}$. In our calculation, we take the value $\delta
t\sim 0.1{\rm s}$ based on the analysis of the BATSE bursts (Shen
\& Song 2003). In lots of other works, $\delta t$ is assumed to be
low to millisecond or even shorter. If that is the case, then the
UV/Optical emission predicted here will be much dimmer unless the
Lorentz factor is much larger. Therefore, the further UVOT
observation can provide the good chance to test our predictions or
impose some important constraints on the poor known parameters of
GRBs.

\section*{Acknowledgments}

Y. Z. Fan thanks Drs. D. Guetta, Z. Li and X. F. Wu for their kind help. We also
thank T. Lu, Z. G. Dai, Y. F. Huang and X. Y. Wang for valuable comments.  We
appreciate the anonymous referees for their useful suggestions that enable us to
improve the paper significantly. This work was supported by the National Natural
Science Foundation of China (grants 10073022,10225314 and 10233010), the
National 973 Project (NKBRSF G19990754).

\end{document}